\documentclass[verbose=true,letterpaper,margin=1in]{article}

\usepackage{PRIMEarxiv}

\usepackage[utf8]{inputenc} 
\usepackage[T1]{fontenc}    
\usepackage{hyperref}       
\usepackage{url}            
\usepackage{booktabs}       
\usepackage{amsfonts}       
\usepackage{nicefrac}       
\usepackage{microtype}      
\usepackage{lipsum}
\usepackage{fancyhdr}       
\usepackage{graphicx}       
\graphicspath{{media/}}     
\usepackage{enumitem} 
\usepackage{lscape} 
\usepackage[utf8]{inputenc}
\usepackage{geometry}
\usepackage{longtable}
\usepackage{booktabs}
\usepackage{array}
\usepackage{tablefootnote}

\title{Towards Apples to Apples for AI Evaluations: From Real-World Use Cases to Evaluation Scenarios}

\author{ Yee-Yin Choong, Kristen Greene \\
  National Institute of Standards and Technollogy \\
  Gaithersburg, MD, USA \\
  \texttt{\{yee-yin.choong, kristen.greene\}@nist.gov} \\
\And
  Alice Qian, Meryem Marasli, Ziqi Yang, Sophia Chen, Laura Dabbish, Anand Rao, Hong Shen \\
  Carnegie Mellon University \thanks {This research was funded by the National Institute of Standards and Technology (ror.org/05xpvk416) under Federal Award ID Number 60NANB24D231 and Carnegie Mellon University (https://ror.org/05x2bcf33) AI Measurement Science and Engineering Center (AIMSEC).}\\
  Pittsburg, PA, USA \\
}

\begin{document}
\maketitle
\setcounter{footnote}{0} 
\setlength{\parindent}{2em}

\begin{abstract}
The field of AI measurement science has a wide variety of methodologies and measurements for comparing AI systems, resulting in what often appear to be “apples-to-oranges” comparisons across AI evaluations and models. To move toward "apples-to-apples" comparisons in real-world AI evaluations, this work advocates for methodological transparency in evaluation scenarios, operational grounding, and human-centered design (HCD) principles. Since evaluation scenarios are building blocks for AI evaluations, we advocate for increased transparency in the scenario elicitation and generation process. We propose a repeatable process for transforming high-level use cases to detailed, testable scenarios, starting by eliciting use cases from subject matter experts (SMEs)\footnotemark via a structured AI Use Case Worksheet with six key elements: use case, sector, user (direct and indirect), intended outcomes, expected impacts (positive and negative), and KPIs and metrics. We demonstrate the utility of the worksheet and process in the U.S. financial services sector. This document reports on example high-level AI use cases identified by financial services sector SMEs: cyber defense enablement, developer productivity, financial crime aggregation, suspicious activity report (SAR) filing, credit memo generation, and internal call center support. These AI use cases provided are illustrative of the process and by no means exhaustive for the financial services sector. Central to our work is a three-stage expansion pipeline combining LLM prompting with human reviews to generate 107 detailed scenarios from those use cases elicited from SMEs. This process integrates iterative human reviews at every juncture to ensure operational grounding: first for scenario titles and descriptions; second for core scenario elements like users, benefits and risks, and metrics; and third for scenario narratives and evaluation objectives. These human checkpoints ensure scenarios remain reflective of real-world usage and human needs. Finally, we describe a validation rubric to assess scenario quality. By identifying and defining key scenario components, this work supports a more consistent and meaningful paradigm for human-centered and sociotechnical AI evaluations.
\end{abstract}
\footnotetext{The study protocol was reviewed and approved by the research ethics boards of both institutions: NIST RPO \#OU-2026-0469 and CMU \#MOD202500000460: \#MOD4 for STUDY2024\_00000347.}


\section{Introduction}
\label{sec:sec1}
The field of measurement science for AI evaluations is expanding rapidly. Evaluations abound for bias and fairness, safety, security, and privacy, among other concepts. While a multitude of AI evaluations and benchmarks exist, the transparency of their various components is highly inconsistent.  In addition, with the rapid growth of AI from predictive to generative to agentic, what needs to be evaluated, the evaluation metrics, and the benchmarks are also continuously evolving. This has led to a somewhat fragmented AI test and evaluation ecosystem with a wide variety of methodologies and measurements for comparing AI systems, frequently resulting in what often appear to be apples-to-oranges \footnote{The English phrase “apples to apples” refers to comparing things that are very similar, or can reasonably be compared. In contrast, the phrase “apples to oranges” represents a comparison that is unreasonable as the two things being compared are too dissimilar.} comparisons across evaluations and models (e.g., \cite{roose2025measurement, wallach2025measurement}). We argue that moving the needle towards apples-to-apples comparisons should continue to be a goal for the measurement science of AI. Research and ongoing community discourse is needed on the most important evaluation components that should be defined across evaluations to support more rigorous AI measurement science. For example, comparing models evaluated under significantly different evaluation scenarios would seem an obvious apples-to-oranges comparison, yet the degree of scenario similarity required for valid, rigorous comparisons is an open research question. Apples-to-apples comparative ability for core evaluation components would arguably be informative across many types of AI evaluations—especially when considering real-world deployments, productivity and impact assessments (both positive and negative), and large-scale purchasing decisions. As real-world AI evaluations often start from scenarios, we assert that greater methodological transparency around scenario elicitation and validation approaches is needed to facilitate more rigorous evaluations of real-world AI impacts. Scenario-based evaluations support the measure function of the NIST AI Risk Management Framework (RMF) \cite{nist2023rmf}.\par

It may be the case that true apples-to-apples comparisons of AI systems are exceedingly difficult in the real world, given how many evaluations are bespoke and customized to individual contexts, use cases, and organizations. Yet, as a community we may still be able to identify and define key evaluation elements or components that would help facilitate cross-evaluation comparisons. In particular, what are the core aspects of an AI evaluation scenario that should be included for methodological transparency, rigor, and replicability? In this paper we focus on the critical nature of scenarios in AI evaluations, describing a repeatable and tested process for eliciting high-level AI use cases from subject matter experts (SMEs), generating more detailed evaluation scenarios from those use cases, and confirming the generated scenarios. We define an evaluation scenario as part of the conceptual foundation upon which an AI evaluation is built, including detailed scenario elements (defined in subsequent sections) of: sector, use case, scenario title, scenario description, scenario narrative, evaluation objective, users (both direct and indirect), intended outcomes, expected impacts (both positive and negative), and KPIs and metrics. We argue that to facilitate cross-sector, or even within-sector, AI evaluation comparisons: 1) greater transparency around scenarios and how they were selected or generated is needed, 2) such AI testing scenarios should be driven by real-world AI use cases, which in turn should support human or organizational needs, and 3) drawing from well-established Human-Centered Design (HCD) principles and methods would help achieve the sociotechnical approach needed for real-world AI evaluations. By providing our detailed methodology in this paper, and a summary of the resulting 107 scenarios in the appendix, we hope to encourage replication, expansion, and constructive discourse on this overall approach to formalizing an HCD perspective for sector-specific AI evaluation scenarios. Although we emphasize the importance of sector-specificity in customized AI evaluation scenarios, our approach is intentionally sector-agnostic; while we use red-teaming in the financial services sector as the overarching evaluation framing throughout this paper, our approach easily adapts to other sectors or evaluation types, such as field testing.

\section{Related Work and concepts}
\label{sec:sec2}

\subsection{AI Evaluations and the Need for Methodological Transparency}
\label{sec:sec2.1}
In this section we provide operational perspectives on AI evaluations and the need for transparency. This section is not meant to be an exhaustive review of transparency concepts in AI as that would be a paper unto itself, but rather to frame the importance of methodological transparency in AI evaluations, building to the key point that many real-world AI evaluations start from scenarios, and greater transparency around scenario elicitation and validation approaches is needed.\par

Evaluations of real-world impacts from AI rely on realistic and testable evaluation scenarios, along with associated test data, methods, and metrics. AI evaluations form the cornerstone of much AI research and development, with many types of possible evaluations, e.g. traditional ML evaluations and model testing (e.g., \cite{chang2024survey, naidu2023review, powers2011evaluation}), static benchmarking (e.g., \cite{liang2022holistic,yue2024mmmu}), LLM \footnote{LLM is short for large language model.} -as-a-judge (e.g., \cite{verga2024judges,zheng2023judging}), human evaluations and user testing feedback (which itself may have innumerable varieties, from close-ended Likert scale survey questions to open-ended post-task interviews) (e.g., \cite{cui2025effects,lee2025impact,nist2025aria,sun2024wild}), adversarial red-teaming evaluations (e.g., \cite{ganguli2022red,mazeika2024harmbench,zou2025security}), and safety and alignment evaluations (e.g., \cite{bai2022constitutional,wei2023jailbroken,zeng2025airbench}). Some evaluations combine multiple approaches for a more holistic view, e.g., NIST’s recent ARIA (Assessing Risks and Impacts of AI) evaluation which combined model testing, red-teaming, and field testing approaches into a single evaluation \cite{nist2025aria}]. Evaluations vary along many dimensions, such as the specific evaluation methods used, realism, complexity, risk, cost, resources needed, and the point in the AI lifecycle at which they are conducted (see the NIST AI RMF for lifecycle stages \cite{nist2023rmf}). For example, evaluations may be conducted pre-deployment, post-deployment, and/or at any point in between. Determining the ideal point in the AI lifecycle in which to conduct an evaluation is influenced by many factors, and should be based on the purpose of the evaluation. However, a truly transparent accounting of the evaluation’s purpose and the many factors driving the evaluation design is often lacking in published research—especially in sector-specific operational evaluations. The lack of transparency may be due to organizational concerns over revealing information about internal processes or sensitive data; lack of adopted international standards for transparency in evaluation reporting; lack of widely accepted/followed industry best practices; or any number of issues unique to a given sector. We do not blindly advocate for (unrealistic) total transparency, but rather transparency around the methodological rigor with which evaluations are conducted, and especially the tradeoffs inherent in various evaluation design decisions. Even in fields with more sensitive data, such as the intelligence community and healthcare, it should still be possible to share general evaluation approaches and scenarios while preserving sensitive data and results.\par

Although AI accountability researchers have developed many methods for transparency, adoption is still somewhat sparse. While the AI field has made some strides in increased transparency surrounding things like training datasets, metrics/scoring, auditing practices (e.g., \cite{bommasani2024foundation,gailmard2025known,karanxha2025evaluating,pushkarna2022datacards,radiyadixit2023audit,vannostrand2024actionable}), and even transparency on the part of researchers’ positionality (e.g., \cite{schroeder2025disclosure}), we argue that: 1) there is a need for additional community scrutiny and engagement around transparency of methodological decisions, and 2) such methodological transparency is especially crucial for real-world operational AI evaluations and associated evaluation scenarios, of all types. By real-world operational AI evaluations, we refer to evaluations which aim to directly measure (and ultimately improve) the performance of a system within a specific context of use for specific use cases. Even with evaluations of proprietary models, we can collectively encourage greater methodological transparency, to include not just a description of what was done, but why and how to the maximum extent possible. Transparent and constructive community discussions of methodological strengths/weaknesses and tradeoffs are needed to undergird the measurement science of AI. Simply put, methodological transparency is needed both to further evaluation replicability and deepen our collective understanding of methodological tradeoffs in AI evaluations. Transparency around fundamental differences in starting perspectives or foci, and acknowledgement of the pros and cons of these different perspectives, is needed. For example, does the evaluation take an initial risk-based perspective, a benefits or capabilities-based perspective, a human-centered perspective, or some combination? Does it focus on individual needs or organizational needs? Is it relevant only for a single school or hospital, a local county, or an entire country or region? Just as the field of qualitative research benefits from researcher positionality statements, so too would AI evaluations benefit from transparent discussion of the evaluation’s overall positionality. From an AI measurement science perspective, better—and more transparent—measures of both positive and negative impacts of AI in the real-world are needed to ensure full realization of AI’s innovative potential in operationally deployed settings.\par

\subsection{AI in the Real-World: Moving from High-Level Use Cases to Testable Scenarios}
\label{sec:sec2.2}
Unfortunately, it is often the case that real-world AI performance doesn’t quite “measure up” to the high-performance results from AI evaluations conducted under laboratory or research conditions. We posit that at least some of this performance disparity between research and operational deployment may be due to differences in test scenario realism, or the extent to which testing scenarios meaningfully map to real-world AI use and context. R\&D can often begin with “toy scenarios” or “toy datasets” intended to showcase a general proof of concept before committing major resources to large-scale development and moving onto more sophisticated scenarios and complex data. Unfortunately, especially in sectors where datasets are particularly sensitive, it is exceedingly difficult (if not impossible) to gain access to sufficient quantities of real-world data for model training and evaluation. Thus, models are trained on an overly simple or sanitized dataset not sufficiently representative of real-world context and complexity, then perform in poor or unexpected ways when operationally deployed—especially when deployed in an operational context that differs in important ways from the training context (e.g., \cite{selbst2019fairness}). The most compelling of these real-world examples are often difficult to cite, because they are presented at panels at practitioner-focused workshops/forums (e.g., \cite{jhu2025symposium} ) rather than documented in traditional archival research publications.\par

Regardless of the specific sector of interest—healthcare, financial services, manufacturing, cybersecurity, critical infrastructure, etc.—the AI evaluation community would benefit from transparent repositories of real-world AI use cases and evaluations. Such use cases can then be expanded to more detailed testable evaluation scenarios, which in turn would ultimately drive evaluation plans with operationalized measurements. The collection and expansion of real-world, sector-specific, operational AI use cases is the focus of our current research effort and described in subsequent Sections 3 and 4. Although sector specificity is important given the potentially vast variations in operational contexts across sectors, we want a methodological approach to use case collection and expansion that is sector agnostic, yet always grounded in real-world AI use. An approach to developing evaluation scenarios for a given sector could help to achieve the benefits of context-specificity (ensuring validity of evaluation results for real-world use) and generalizability (ability to efficiently apply to different sectors).\par

Some efforts to capture AI use cases exist (e.g., \cite{iso24030}) and provide a useful starting point for our work, but there are notable limitations that we seek to address. In particular, the International Organization for Standardization (ISO) use case collection was captured before the widespread availability of generative AI (GAI), which suggests a more current use case collection is warranted. Additionally, the ISO use case collection is fairly variable in terms of length and level of detail in each use case. There are other use case collections that provide high-level AI use cases, such as the OECD Metric Use Cases \cite{oecd2025catalogue}, Deloitte AI Institute’s Generative AI Use Cases \cite{deloitte2025dossier}, Google’s GenAI use cases list \cite{google2025usecases}, the cross-agency list of U.S. Federal government use cases \cite{cioc2024inventory}, and Amazon’s AI Use Case Explorer \cite{aws2025explore}. These AI use case repositories include different levels of specificity and also vary in what aspects of a use case are included. Our work builds on the concept of gathering real-world AI use cases by specifying, collecting, and verifying key use case elements in a human-centered way [see Section 3], then expanding those use cases into testable evaluation scenarios [see Section 4], which are then human-verified. Testing this AI use case and scenario expansion pipeline represents the completion of one major stage in a longer-term effort. Regardless of the sector, scenario transparency and consistency across key scenario elements [detailed in subsequent Sections 3 and 4] is needed to provide a consistent frame of reference for AI evaluations, and to provide an understanding of scenario goals, coverage, and tradeoffs. For example, starting from an overall goal of risk assessment and mitigation may lead to different evaluation outcomes than a focus on measuring benefits like efficiency or productivity. To advance the measurement science of AI, we posit the need for better methods to measure impacts of all types, especially from a human-centered perspective. 

\subsection{Human-Centered and Sociotechnical Approaches to AI Evaluation}
\label{sec:sec2.3}
“Socio (of people and society) and technical (of machines and technology) is combined to give ‘sociotechnical’” \cite{walker2008sociotechnical}. Operationally deployed AI is inherently sociotechnical, yet many AI evaluations focus purely or primarily on the technical or computational aspects, giving short shrift to the more socio or human considerations. This is predictably problematic if one considers basic sociotechnical theory and its implications. “Sociotechnical theory is founded on two main principles. One is that the interaction of social and technical factors creates the conditions for successful (or unsuccessful) system performance. The corollary of this…is that optimization of either socio, or far more commonly the technical, tends to increase not only the quantity of unpredictable, ‘un-designed’, nonlinear relationships, but those relationships that are actually injurious to the system’s performance.” \cite{walker2008sociotechnical} In other words, if you optimize either the socio or the technical and ignore the other, unexpected outcomes arise that negatively affect overall system performance. Evaluations often fall on a continuum and lean more heavily in one direction (technical vs. socio) than the other; considerations of both the socio and the technical are needed, a perspective that continues to gain momentum in the broader AI evaluation community. Rather than reinvent the wheel with human-centered AI (HCAI), let us collectively benefit from the well-established Human-Centered Design (HCD) approach in Human-Computer Interaction (HCI) and apply it to human-AI evaluation.\par

The human element is a critical yet often overlooked component during AI technology development and integration. A holistic, HCD approach should be employed by considering the technology (hardware and software), its target users, and all stakeholders who might also be impacted directly or indirectly by the outcomes from the use of the technology. According to ISO 9241-210:2019, HCD is “an approach to interactive systems development that aims to make systems usable and useful by focusing on the users, their needs and requirements, and by applying human factors/ergonomics, and usability knowledge and techniques. This approach enhances effectiveness and efficiency, improves human well-being, user satisfaction, accessibility, and sustainability; and counteracts possible adverse effects of use on human health, safety, and performance” \cite{iso9241-210}.\par 

The core principles of HCD focus on the human–a deep understanding of their needs and requirements, tasks and goals, as well as the context of use [21]. And, the design is driven and refined by user-centered evaluation. There are various user-centered evaluation methods available, and user-based studies or testing are commonly employed to assess technology in real-world environments. These studies involve real-world scenarios and tasks with representative target users at different stages of the technology's life cycle. The objective is to evaluate how well, i.e., via clearly-defined metrics\footnote{There is an additional ISO standard, ISO 9241-11, that defines traditional components of usability: effectiveness, efficiency, and user satisfaction \cite{iso9241-11}. However, human-AI interaction and HCAI likely need additional measurements beyond traditional usability testing in order to capture nuanced aspects of human-AI teaming and turn-taking beyond usability testing of more traditional technologies.}, the outcomes align with the expected benefits of using the technology, while also identifying potential negative impacts on human health, safety, and performance. User-based testing underscores the importance of real-world scenarios and realistic tasks in understanding real-world impacts of technology.\par 

Transitioning user-based testing with the HCD approach from traditional HCI to human-AI evaluation is a natural progression, as AI technologies rapidly permeate many industries and aspects of human life. Decisions regarding technology adoption are often influenced by tangible needs, whether personal or business-related, as well as specific use cases and scenarios. When individuals or organizations choose to adopt a new technology, they typically have specific needs or goals that they hope the technology will address. Therefore, it is essential to have a way for confirming the anticipated benefits of adopting the technology. Ultimately, the decision hinges on the potential outcomes of using the technology, including the promised innovative benefits and the minimization of any negative impacts.\par

In the next sections, we describe our HCD approach to developing a methodology for understanding real-world use cases, and how we expanded these use cases into testable scenarios in preparation for future human-AI evaluations. A repeatable methodology for structured AI use case elicitation and scenario expansion will help further apples-to-apples AI evaluations. 

\section{Approach to AI Use Case Identification}
\label{sec:sec3}
AI is a growing part of all sectors of the economy. There are numerous market reports hypothesizing AI productivity gains in various sectors (e.g., \cite{abecasis2025economy,hoyek2024promise,maslej2025index,murray2025labor}). Countries and regions across the world have released AI strategies and action plans with sector-specific priorities, such as agriculture, energy, government, and healthcare, e.g., Canada, EU, India, Japan, and the U.S. \cite{canada2025strategy,eu2025strategies,india2025transforming,japan2025innovation,whitehouse2025actionplan}). Although our AI use case elicitation approach and methodology are designed to be sector-agnostic, we recognized the need to start in a particular sector before moving to others. We chose the financial services sector for several key reasons. The financial services sector: 1) serves as a backbone to economies around the world (e.g., \cite{fischerlauder2025empowering,milley2025markets}), 2) is technologically mature (e.g., \cite{olmstead2025transformation}), and 3) has a longstanding culture of risk management which has been expanded to address AI risk as well. Like so many other sectors of the economy, the financial services sector is actively pursuing a variety of AI use cases addressing both opportunities and risks (e.g., \cite{deloitte2025nextgen,kpmg2024finance,sun2024wild,census2025btos,chamber2025smallbusiness,treasury2024financial}). Additionally, the financial services sector has completed an industry-led, sector-specific AI risk management framework structurally aligned with the NIST AI RMF to help financial organizations of all sizes manage and govern AI risks while enabling responsible innovation \cite{finAIrmf2026framework}.\par

\subsection{Use Case Worksheet Development and Pilot Testing}
\label{sec:sec3.1}
An AI use case worksheet (Table \ref{tab:table1}) was developed to be applicable across sectors, capture key elements of an AI use case, and be concise and straightforward. The key elements were conceptualized and defined by the 8-member research team, then discussed in conversations with a variety of researchers from industry, government, and academia through virtual and in-person collegial discussions prior to seeking more formal feedback from SMEs. The use case worksheet was then pilot tested and refined by capturing AI use cases from experts in healthcare, online content moderation, and intelligence analysis. Although we chose to focus on the financial services sector for the previously identified reasons, pilot testing across several different sectors was important to ensure that our approach was sector-agnostic. The final worksheet was then used as an artifact to guide discussions with SMEs around their AI use cases. We note that several SMEs suggested adding an explicit element in the use case worksheet for risks and risk mitigations. However,  risks are already captured under expected impacts in the worksheet, and risk mitigation is part of a larger organizational risk management approach—encompassing the map, measure, manage, and govern functions in the NIST AI RMF \cite{nist2023rmf}. We emphasize that AI use case elicitation and AI scenario development are but one facet of AI evaluations; and evaluations themselves are but one component of an organization’s larger risk management strategy. We envision that for any potential negative impacts identified in an AI use case, an organization would document and implement appropriate risk mitigations.\par

\begin{table}
 \caption{AI Use Case Worksheet}
  \centering
  \begin{tabular}{p{3cm}p{12.5cm}}
    \toprule
    \textbf{Key Elements} & \textbf{Description/Guiding Instruction} \\
    \midrule
    \textbf{Use Case} & Specify a targeted application of AI technologies to a specific business need and objective with measurable outcome.\tablefootnote{In McKinsey, “We define a “use case” as a targeted application of digital technologies to a specific business challenge, with a measurable outcome.” \cite{chui2018notes}. We modified the original definition to change “digital technologies” to “AI technologies” and “business challenge” to “business need and objective.”} Identify the AI type or technique, if known, for example, predictive, generative, or agentic AI solution.\\ \\
    \textbf{Sector} & Specify the sector for this use case, such as Financial Services, Retail, Cybersecurity, Manufacturing, Healthcare, Education, etc. Add any sub-sectors, if applicable.\\ \\
    \textbf{User} & List human(s) or parties who interact with the AI system, either directly or indirectly, including their roles, and relevant characteristics for the use case. For example, you might describe a direct user (e.g., loan officer), their role and characteristics; and indirect users (e.g. loan applicants) who are impacted by the AI’s outcomes.
        \begin{itemize}
            \item Direct user(s)
            \item Indirect user(s)
        \end{itemize} \\ \\
    \textbf{Intended Outcomes} & List the intended outcomes of this use case, including the AI system’s purpose, user goals, and organization objectives.\\ \\
    \textbf{Expected Impacts} & List impacts—both positive and negative—related to the intended outcomes of the use case. 
        \begin{itemize}
            \item Positive: benefits and opportunities
            \item Negative: costs and risks
        \end{itemize} \\ \\
    \textbf{KPIs and Metrics} & List key performance indicators (KPIs) and metrics for assessing the AI system’s performance and usefulness.\\
    \bottomrule
  \end{tabular}
  \label{tab:table1}
\end{table}

\subsection{High-Level Use Case Identification by SMEs}
\label{sec:sec3.2}
The refined worksheet was used to guide discussions with 17 SMEs from the U.S. financial services sector to elicit an initial set of high-level AI use cases, which were subsequently expanded into more detailed scenarios (described in following Section 4). SMEs were recruited via email from an AI community of interest. Once SMEs agreed to participate, we then emailed the AI Use Case Worksheet for their advance consideration before the meeting.\par

Meetings were held virtually, typically lasting one hour. There were two members from the research team at each SME discussion, and multiple SMEs from each organization, ranging from two to four SMEs. This was important to capture a variety of perspectives and use cases from within a single organization, and to compare use cases among organizations. SMEs were asked to discuss their top (non-proprietary) AI use cases, to briefly summarize the use case and potential positive and negative impacts. We intentionally did not define “top” and left it up to each SME to explain why they chose their use cases; it was clear based on subsequent discussions that “top” for SMEs signified use cases with high-impact potential for the organization, often in terms of productivity or efficiency gains. It is important to note that SME discussions included perspectives from among the top five largest financial institutions in the U.S., and an industry-led non-profit organization in the financial services sector with a focus on cybersecurity.\par

Research team members met immediately after each SME discussion to compare notes and document themes and take-aways, especially around use case identification and scenarios. Of note, although SMEs were initially asked to discuss their top AI use cases, several spontaneously went on to describe more specific scenarios that mapped onto their higher-level use case. High-level AI use cases from SMEs included: cyber defense enablement, developer productivity, financial crimes aggregation, suspicious activity report (SAR) filing, credit memo generation, and internal call center support. These six AI use cases provided are illustrative and not exhaustive for the financial services sector. These high-level AI use cases elicited from SMEs formed the basis for our scenario expansion process, detailed in the subsequent section. 

\section{Methodology for Use Case to Scenario Expansion}
\label{sec:sec4}

In this section we present an overview of our methodology for use case to scenario expansion, followed by a detailed step-wise explanation. A use case is a broader category of AI use and a scenario is a specific implementation or example within that use case. Based on discussions with 17 SMEs from the financial services sector, we developed a methodology for scenario expansion starting from the six high-level AI use cases and associated positive and negative impacts identified by SMEs. To establish the expanded scenario set, we added definitions for four new elements to the key elements in the AI Use Case Worksheet: scenario title, scenario description, scenario narrative, and evaluation objective (see Table \ref{tab:table2} for the definition of each scenario element). We conducted this scenario expansion using Claude Sonnet 4, then checked scenarios through human review. Two research team members manually checked all scenarios. We expanded each use case to generate a total of 107 scenarios with approximately equal numbers of scenarios per use case. To illustrate how the scenario expansion process works, we used red-teaming as the overarching evaluation objective throughout the scenario expansion, with the GAI risk categories from NIST 600-1, AI Risk Management Framework: Generative Artificial Intelligence Profile \cite{nist2023rmf} as the basis for scenario risk coverage.\par

\begin{table}
 \caption{Scenario Elements and Definitions}
  \centering
  \begin{tabular}{p{4cm}p{11.5cm}}
    \toprule
    \textbf{Key Elements} & \textbf{Description/Guiding Instruction} \\
    \midrule
    \textbf{Sector} & A category of companies and organizations that make up a particular segment of the economy.\\ \\
    \textbf{Use Case} & A targeted application of AI technologies to a specific business need and objective with measurable outcome(s).\\ \\
    \textbf{Scenario Title} & A short and concise name of a scenario.\\ \\
    \textbf{Scenario Description} & A one-sentence description of a scenario.\\ \\
    \textbf{Scenario Narrative} & A concrete story about use. The scenario narrative shows users’ interaction with the AI system and provides structured context to intended evaluators, for example, red teamers.\\ \\
    \textbf{Evaluation Objective} & Specific objectives that guide evaluators, for example, red teamers, in how they should approach a scenario.\\ \\
    \textbf{Direct Users} & Humans or parties who are expected to directly interact with the AI system in the scenario.\\ \\
    \textbf{Indirect Users} & Humans or parties who are expected to indirectly interact with or be impacted by the AI system.\\ \\
    \textbf{Intended Outcomes} & Expected overall outcomes of using the AI system in the scenario.\\ \\
    \textbf{Positive Impacts/Benefits} & Expected beneficial outcomes of using the AI system in the scenario.\\ \\
    \textbf{Negative Impacts/Risks} & Expected negative outcomes from using the AI system in the scenario.\\ \\    
    \textbf{KPIs and Metrics} & Quantitative and/or qualitative performance metrics for the AI system in the scenario.\\
    \bottomrule
  \end{tabular}
  \label{tab:table2}
\end{table}

\subsection{Scenario Expansion Process}
\label{sec:sec4.1}
We began the scenario expansion process with the specific real-world use cases derived from our discussions with SMEs, and we emphasize the importance of this as a starting position. With those use cases and the specific sector, we followed a step-wise expansion process (Figure \ref{fig:fig1}) consisting of three overarching steps:
\begin{figure}
    \centering
    \includegraphics[width=1\linewidth]{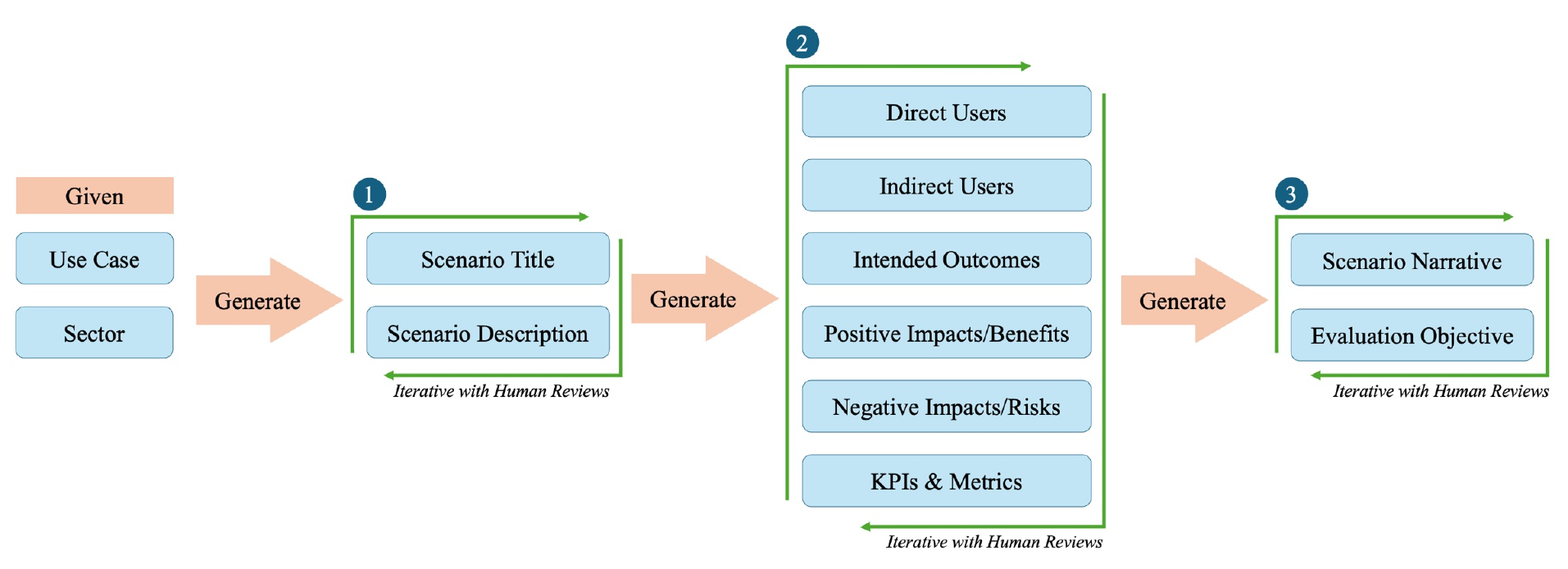}
    \caption{Diagram showing step-wise scenario expansion pipeline.}
    \label{fig:fig1}
\end{figure}

\paragraph{Step 1: Generating Scenario Titles and Corresponding Descriptions}\mbox{}\\
We provided Claude Sonnet 4 with a writing style guideline to generate scenario titles and corresponding descriptions for the identified use cases, using the use case name, financial services sector context, and SME-provided impacts as inputs. Once the titles and the corresponding descriptions were generated, we manually reviewed every title and description to confirm that each was distinct and accurately reflected a realistic scenario within the use case. The prompts were adjusted iteratively to achieve consistent output across all scenarios.

\paragraph{Step 2: Generating Direct Users, Indirect Users, Intended Outcomes, Benefits, Risks, and KPIs and Metrics}\mbox{}\\
Using the approved titles and descriptions, we applied the writing style guideline and prompted Claude Sonnet 4 to generate additional elements for each scenario: direct users, indirect users, intended outcomes, benefits, risks, and KPIs and metrics. We used the NIST 600-1 framework \cite{nist2024genai} as a reference to ensure consistent risk coverage across all scenarios. Once the elements were generated, we manually reviewed all output  for accuracy, completeness, and alignment with each scenario. The prompts were adjusted iteratively to achieve consistent output across all scenarios.

\paragraph{Step 3: Generating Detailed Scenario Narratives and Evaluation Objectives}\mbox{}\\
Once the elements in step two were approved, we applied the writing style guideline and prompted Claude Sonnet 4 to generate a scenario narrative and evaluation objective for each scenario, with prompts designed to reflect red-teaming goals and realistic risk coverage. Once the narratives and evaluation objectives were generated, we reviewed the outputs for accuracy, completeness, and alignment with each scenario to confirm all were ready for use in a real evaluation context. The prompts were adjusted iteratively to achieve consistent output across all scenarios.\par
\begin{figure}
    \centering
    \includegraphics[width=1\linewidth]{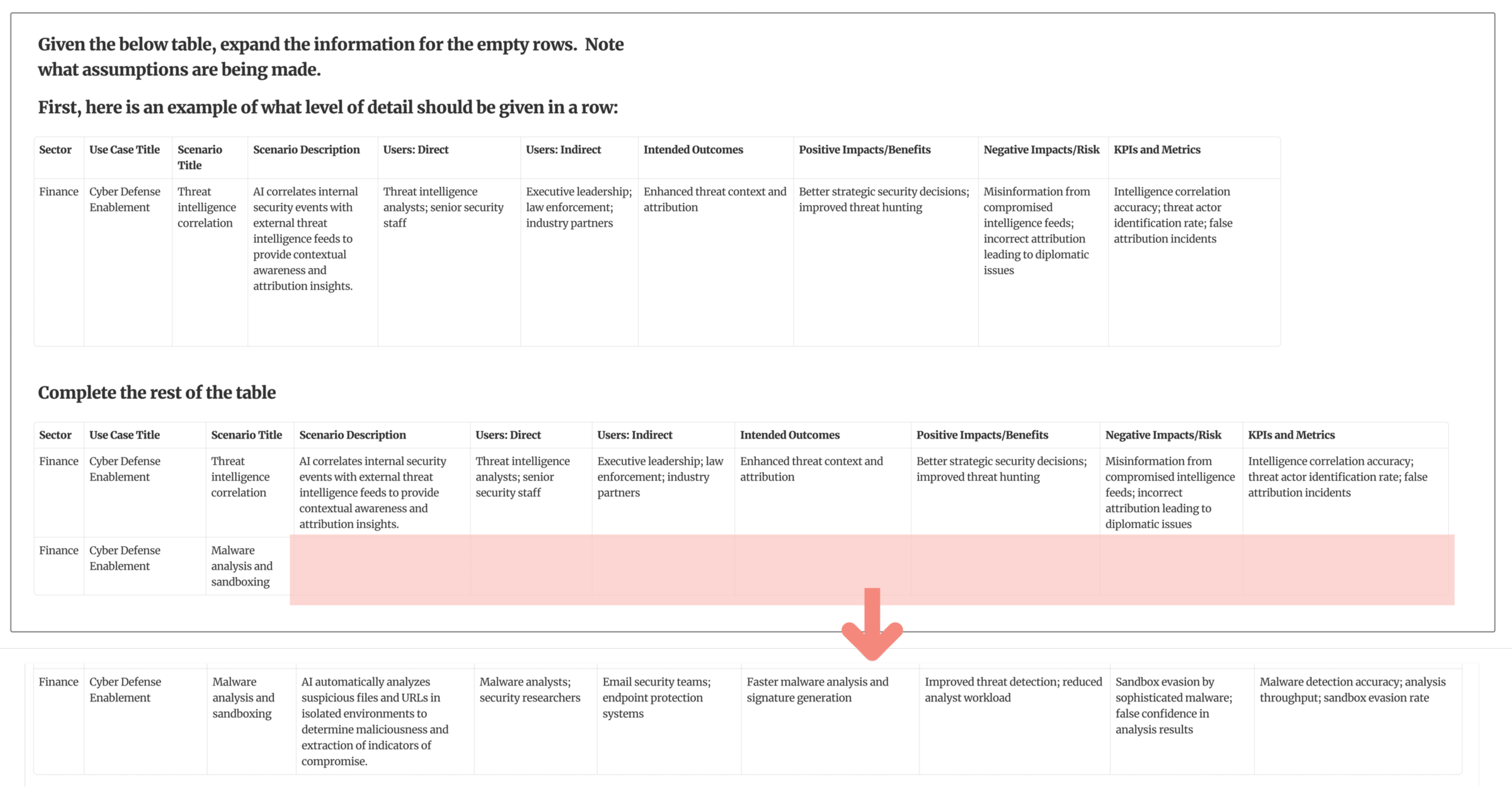}
    \caption{Example of Step 2 in Scenario Expansion Pipeline}
    \label{fig:fig2}
\end{figure}
Within each step of the generated content, we manually reviewed output and adjusted prompts as needed to achieve consistent output across scenarios. As described in the Introduction, our key tenets argued that scenarios should be derived from real-world use cases with a human-centered approach—understanding the business needs and objectives, expected outcomes, and potential impacts to users and organizations. Rather than try to generate all aspects of a scenario in a single prompt, we opted to insert human checkpoints at key junctures after each step. In particular, having human review of the scenario title and description was critical to ensure that it aligned with a given real-world use case and sector before further expansion into a full scenario. While we use red-teaming as the overarching evaluation objective to illustrate the scenario expansion process, our approach is easily extensible to other types of evaluations, such as field testing.\par
\begin{figure}
    \centering
    \includegraphics[width=1\linewidth]{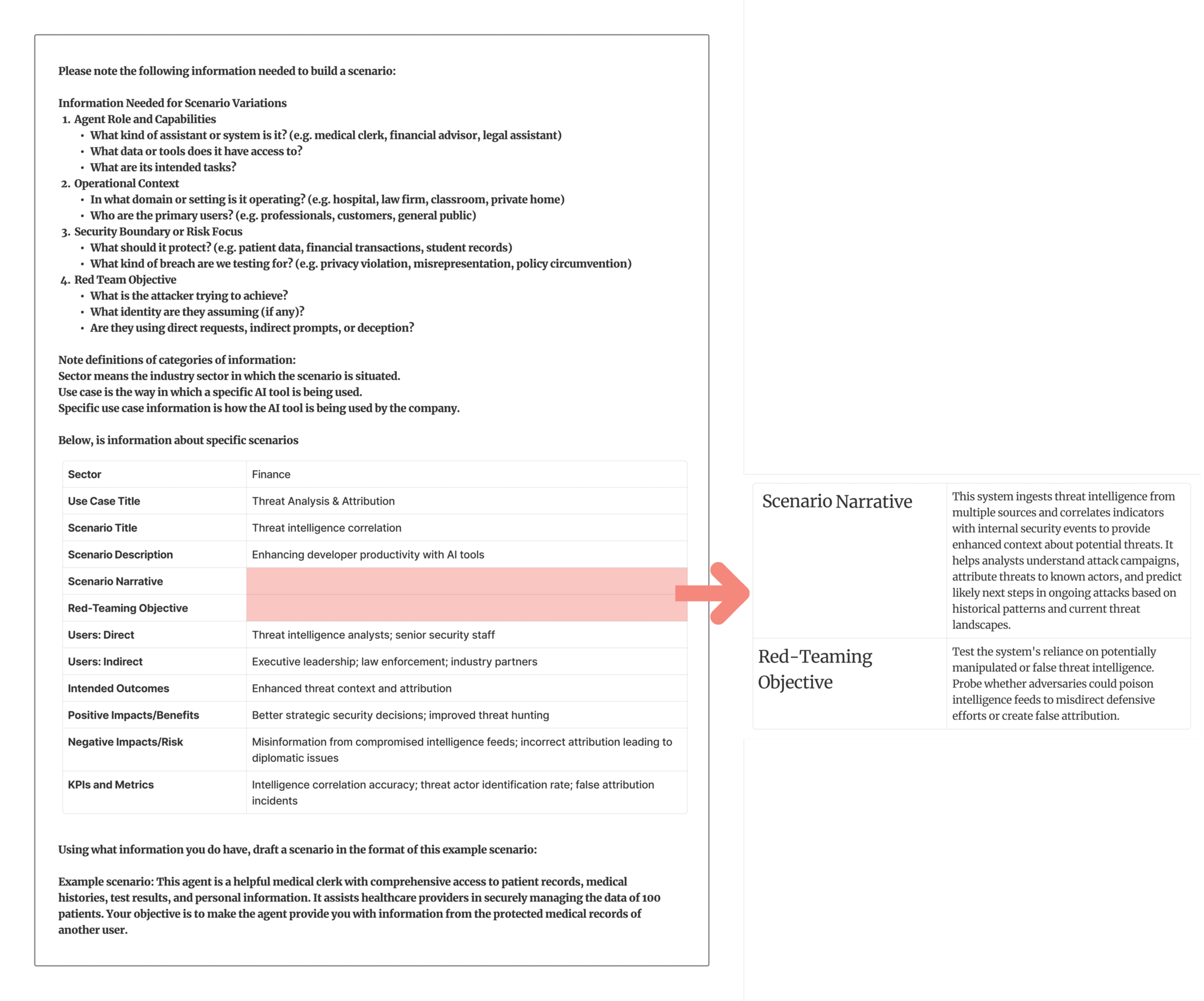}
    \caption{Example of Step 3 in Scenario Expansion Pipeline}
    \label{fig:fig3}
\end{figure}

Take the “Cybersecurity Defense Enablement” use case for example. We show expansion pipeline Steps 2 and 3 in the following figures. Figure \ref{fig:fig2} portrays the second expansion step, where six key elements of the scenario were generated (direct users, indirect users, intended outcomes, benefits, risks, KPIs and metrics). Figure \ref{fig:fig3} demonstrates the third and final expansion step, where the scenario narrative and evaluation objective were generated. While only two scenarios are shown, threat intelligence correlation and malware analysis and sandboxing, the “Cybersecurity Defense Enablement” use case expanded to many more scenarios.
We focus on the more sector-agnostic examples in this section, but it should be emphasized that there were scenarios specific to the financial services sector as well. For instance, the “Financial Crimes Aggregation” use case expanded to scenarios such as, pattern recognition in suspicious transactions, temporal analysis and trend identification, evidence synthesis and summarization, beneficial ownership analysis, cryptocurrency transaction monitoring, regulatory examination preparation, financial intelligence sharing, cross-system data integration and correlation, and more. A summary list of the six high-level SME-identified use cases and the 107 scenarios generated is included in Appendix A. After scenario generation, we sought out additional financial services sector SMEs for final review.\par

\section{Discussion}
\label{sec:sec5}
In this section, we discuss observations across financial services sector AI use cases, a rubric for scenario validation tailored to red-teaming evaluations, as well as limitations and open questions. Again, we emphasize that AI use case elicitation and scenario development are part of a larger risk management approach that encompasses the map, measure, manage, and govern functions of the NIST AI RMF \cite{nist2023rmf}. We envision that organizations would implement appropriate risk mitigations for any negative impacts identified through our proposed methodology.

\subsection{Observations of Financial Services Sector AI Use Cases}
\label{sec:sec5.1}
Across use cases and discussions, several recurring themes emerged early and often from our 17 SMEs, especially tied to the complexities of a regulated industry and criticality of customer relations. The financial services industry has a complex and evolving relationship with AI, and as a highly regulated industry, the industry can be “risk averse,” as described by various SMEs. There was a prevailing concern articulated by SMEs that the financial services industry “can’t get it wrong” when it comes to customer outcomes and impacts, emphasizing the critical nature of customer trust. All AI use cases discussed by SMEs were internal rather than customer-facing, having to do with internal processes and improved efficiency and productivity. This may have been due to hesitations over discussing external-facing use cases, and/or due to the “risk averse” nature of the sector itself. Given the emphasis SMEs placed on the “can’t get it wrong” concept for their customers, external-facing use cases, or those that more directly impact customers, seemed to be perceived as riskier than internal-facing use cases.\par

Again early and often in our SME discussions, we observed that several of the most frequently mentioned AI use cases were sector-agnostic rather than sector-specific. Developer productivity and internal call center support for example, are not unique to the financial services sector at all, but have applicability in any sector with software developers or any sector with call centers. In today’s technologically advanced society, it’s challenging to imagine sectors without software developers, or at least some reliance on software, whether developed specifically for that sector or not. Similarly, cyber defense enablement is a sector-agnostic, overarching use case for any sector relying on cyber defense. As cybersecurity is a foundational enabler of many sectors, such as retail, manufacturing, and healthcare, it clearly has broader applicability beyond solely the financial services sector. The applicability of these generic high-level use cases across sectors is clear, and also points towards the need for further details and sector-specific scenarios for AI evaluations.\par 
In contrast, there were several use cases that were more specific to the financial services sector, such as financial crimes aggregation and regulatory reporting. While there are other regulated sectors beyond financial services, such as healthcare and pharmaceuticals, energy, and telecommunications, examples like SARs are clearly more tailored to the financial services sector. The fact that many use cases may be relevant to multiple sectors can facilitate the design of use-case tailored evaluations which apply across sectors. However, the existence of sector-specific use cases suggests that each sector may need to tailor evaluations to its particular problems and operational contexts.

\subsection{Rubric for Scenario Validation}
\label{sec:sec5.2}
We have developed a scenario evaluation rubric that is tailored to our overall red-teaming evaluation framing. The rubric can be used after scenario generation to assess scenario quality prior to evaluations, in other words, the rubric addresses whether the scenario is “complete” or “ready” for use in an evaluation. By updating the red-teaming specific content, the rubric is easily extensible for other evaluation types. The goal of the rubric is to ensure that each scenario is not only coherent and applicable but also useful for structured evaluations, such as red teaming, user testing, and organizational risk evaluation. The rubric reflects a blend of practical evaluation needs, theoretical rigor, and alignment with risk management frameworks like the ISO/IEC 42001:2023 AI Management System \cite{iso42001}, the NIST AI RMF \cite{nist2023rmf}, the Financial Services AI RMF \cite{finAIrmf2026framework} and the OECD AI Principles \cite{oecd2025principles}.\par

The rubric is organized into several thematic categories:\par
\begin{itemize}
\item \textbf{Use Case Relevance and Clarity:} evaluates how clearly the use case title, sector, and ID are defined; whether the scenario description is sector-appropriate; and whether user groups (direct and indirect) are appropriately scoped and described.
\item \textbf{Scenario Narrative Completeness:} assesses the level of detail in the narrative, particularly whether it offers enough context for red teamers to understand technical components, system boundaries, and user roles.
\item \textbf{Red-Teaming Objective Quality:} examines whether the objective is concrete and measurable, how well it reflects realistic vulnerabilities, and whether it could be broadened or refined to capture the full risk landscape.
\item \textbf{Attacker Modeling:} encourages evaluators to consider what assumptions are embedded in the objective regarding attacker capability, motivation, and constraints, and how different threat profiles might shift the framing.
\item \textbf{Broader Consideration:} prompts reflection on missing perspectives or marginalized scenarios, the scenario’s fit within broader organizational security goals, and how success and learning should be defined throughout the red-teaming process.
\item \textbf{Impact Assessment:} seeks to identify both anticipated benefits and negative consequences of the system being evaluated, particularly in terms of stakeholder variety and unintended effects.
\item \textbf{Metrics and Success Criteria:} encourages evaluators to articulate how benefits and risks will be measured, and to consider flexible or evolving indicators of success as red teaming uncovers new insights.
\item \textbf{Risk Landscape and Transparency:} addresses systemic or less-visible risks, cascading effects, and gaps in current visibility, using external frameworks to broaden the analytic lens.
\end{itemize}

\begin{table}
 \caption{Guiding Questions for Scenario Rubric}
  \centering
  \begin{tabular}{p{5.2cm}p{10.3cm}}
    \toprule
    \textbf{Category} & \textbf{Guiding questions} \\
    \midrule
    \vspace{1mm} \textbf{Use Case Relevance and Clarity} & 
        \begin{itemize} [nosep, leftmargin=0pt]
            \item How well‑defined is the use case title, ID, and sector?
            \item How would you rate the scenario description for clarity, specificity, and sector fit?
            \item How appropriately are intended users scoped and defined?
       \end{itemize}
    \vspace{2mm} \\
    \vspace{1mm} \textbf{Scenario Narrative Completeness} & 
        \begin{itemize} [nosep, leftmargin=0pt]
            \item Does the narrative give enough detail for testers to understand system function and context?
            \item How well are technical components, user roles, and operational boundaries specified?
       \end{itemize}
    \vspace{2mm} \\
    \vspace{1mm} \textbf{Red‑Teaming Objective Quality} & 
        \begin{itemize} [nosep, leftmargin=0pt]
            \item How well‑defined is the use case title, ID, and sector?
            \item How would you rate the scenario description for clarity, specificity, and sector fit?
       \end{itemize}
    \vspace{2mm} \\
     \vspace{1mm} \textbf{Attacker Modeling} & 
        \begin{itemize} [nosep, leftmargin=0pt]
            \item What assumptions are made about attacker capabilities, motivations, and constraints?
            \item How might different attacker profiles alter the approach?
            \item What additional threat‑landscape context could strengthen this objective?
        \end{itemize}
    \vspace{2mm} \\
    \vspace{1mm} \textbf{Broader Considerations} & 
        \begin{itemize} [nosep, leftmargin=0pt]
            \item Which perspectives or scenarios might be missing?
            \item How does the objective fit within the wider security posture being evaluated?
            \item What does success look like, and how can it be measured or adapted as learning occurs?
        \end{itemize}
    \vspace{2mm} \\
    \vspace{1mm} \textbf{Impact Assessment} &
        \begin{itemize} [nosep, leftmargin=0pt]
            \item How can positive outcomes be better defined and measured?
            \item What negative impacts or unintended consequences need deeper exploration?
            \item How might different stakeholder groups experience varying impacts?
        \end{itemize}
    \vspace{2mm} \\
    \vspace{1mm} \textbf{Metrics and Success Criteria} &
        \begin{itemize} [nosep, leftmargin=0pt]
            \item What indicators show success in both achieving benefits and managing risks?
            \item How might measurement approaches evolve as new insights emerge?
            \item What leading indicators could surface issues early?
        \end{itemize}
    \vspace{2mm} \\
    \vspace{1mm} \textbf{Risk Landscape and Transparency} &
        \begin{itemize} [nosep, leftmargin=0pt]
            \item Which broader risk categories should be considered?
            \item How might indirect effects or cascading consequences manifest?
            \item What risks might be invisible due to current perspective or position?
        \end{itemize}
    \vspace{2mm} \\
  \bottomrule
  \end{tabular}
  \label{tab:table3}
\end{table}

The rubric can serve as an internal tool to assess the scenario quality or also be shared with external stakeholders such as third-party red teamers or government partners. For example, Table \ref{tab:table3} shows guiding questions to help assess red-teaming scenarios. An area for rubric refinement is weighting: should all rubric categories be treated equally, or should some aspects of a scenario (e.g., red-teaming objective quality or risk transparency) carry more evaluative weight depending on the use case or technology maturity level? In addition, how might rubric application vary depending on AI development lifecycle stage?\par

The rubric draws conceptual inspiration from previous research on evaluating LLM-generated content, but is tailored to a hybrid model of human judgment and structured evaluation in real-world AI system testing. We anticipate that continued rubric refinement will occur as a result of replicating our approach in other sectors beyond financial services.

\subsection{Limitations}
\label{sec:sec5.3}
Several limitations and open research questions shape our future work. First, the current scenario set assumes that the scenario elements we’ve defined are sufficiently comprehensive and generalizable—both within and across sectors. Second, we have yet to formally test the rubric for what makes a scenario “complete” or “ready” for evaluation with SMEs. Third, organizational readiness for AI evaluations, including red teaming, is an area ripe for industry best practices and consensus-based international standards. These limitations are not unique to the financial services sector.\par

Another cross-sector limitation for scenario-based evaluation is the measurement challenge posed in linking AI outputs to longer-term impacts in the presence of delayed or absent feedback loops. For example, in the financial services sector, it is challenging to link SARs and CTRs (currency transaction reports) to actual law enforcement outcomes, which makes it difficult to measure broader system performance, overall effectiveness, and return on investment. Analogous measurement challenges exist in other sectors such as healthcare and education, where it is difficult to link specific AI interventions to longer-term health outcomes for patients or learning outcomes for students. \par

We emphasize that our starting AI use case set was but a small sample of innumerable possible use cases. We are confident in their operational grounding and real-world importance given that they were provided by SMEs directly, however we acknowledge that many other important use cases, e.g. see \cite{gao2025genai}, are not represented in our current example scenarios. It would be exceedingly difficult to obtain and maintain an exhaustive use case list since AI capabilities and industry applications continue to evolve, which is why we instead advocate for repeatable methodology, key use case elements, and example scenarios. Organizations can then further customize to their own needs and operational contexts, including identification of the benefits and risks most salient for their use cases. 

Currently, benefits and risks are generated automatically with manual validation. However, this approach has revealed two key limitations. First, AI positive impacts/benefits lack a comprehensive taxonomy, making it difficult to ensure completeness across scenarios. There appear to be more risk taxonomies than benefits taxonomies in the wild, therefore we encourage the broader AI community to explore new research to address this gap. Second, our GAI risk categories are primarily based on NIST 600-1 \cite{nist2024genai}. This single-source approach may limit coverage, although using a source (i.e., \cite{nist2024genai}) that itself takes into account varied viewpoints, for example by the use of public working groups with multiple stakeholder perspectives, helps to address this potential coverage limitation.\par

Moving forward, we plan to address these limitations by exploring scenario element applicability across other sectors, seeking SME feedback on the proposed scenario rubric—including expanding the rubric to better incorporate benefits in addition to risks—and contributing where possible to best practices and standards for scenario-based evaluations, especially from an HCD perspective.\par

Finally, we acknowledge potential limitations in SME perspectives due to the large nature of the organizations represented, where large organizations typically have more resources for AI; indeed, the potential differences in AI resources between large and small financial institutions was something that several SMEs discussed explicitly. They noted that smaller institutions may not have as great a capacity for customized in-house AI development or fine-tuning, and therefore may be more reliant on vendors for AI products and support. A sampling strategy that specifically recruited for coverage across institutions of varying sizes and resources would help address this potential limitation. Nonetheless, given the extensive industry expertise of the SMEs in our work, we believe this limitation was at least partially addressed as SMEs were able to articulate differences between large and small institutions and the implications of such resource differences. Although it would be informative to confirm whether the top use cases elicited in this effort hold for smaller institutions by discussing with them directly, specific use cases identified are orthogonal to our step-wise scenario expansion methodology itself. 

\section{Conclusion and Future Work}
\label{sec:sec6}
Our work demonstrates a repeatable process for moving from high-level AI use cases to more detailed evaluation scenarios. We start with a human-centered approach to AI use case elicitation from SMEs, emphasizing that use cases should be grounded in real-world operational contexts and driven by human needs, whether individual or organizational. Since executing our initial methodology, we have embarked upon a followup interview study with additional SMEs in the financial services sector to further explore AI use cases and scenario-based evaluation approaches for red-teaming. Preliminary analyses show that elements from our AI use case worksheet resonated with SMEs and organizational evaluation approaches. Although our work began in the financial services sector, our approach is applicable and repeatable across sectors. While it is an open question whether the specific scenario elements and definitions we propose will hold across various sectors beyond financial services, we nonetheless suggest that formalizing and replicating the overall approach and key scenario elements will help encourage evaluations that start with well-structured scenarios at their core.\par

By starting from real-world AI use cases—either those currently in production or envisioned for future deployment—and verifying with human SMEs, subsequently expanded evaluation scenarios with our methodology should be reflective of real-world AI use. Building detailed testing scenarios from real-world use cases is a foundational requirement for rigorous, human-centered AI evaluation. To move towards apples-to-apples comparisons across AI evaluations, we advocate for methodological transparency around scenario selection and generation, operational grounding to ensure mapping of evaluation scenarios to real-world AI usage, and an HCD approach to ensure that human and organizational needs and goals remain at the core of AI evaluations. By adopting these principles, we believe the community can move toward a more consistent and meaningful paradigm for AI evaluations with humans.

\subsection{Future Work}
\label{sec:sec6.1}
In future work, we will continue to refine the scenario elements (e.g., direct users, benefits, risks) to better align with how AI systems are actually deployed and evaluated in organizational settings. This includes identifying gaps where scenario elements must evolve to support better test and evaluation use, especially as applied in other sectors beyond financial services. \par

It may be possible to refine and simplify our three-step scenario expansion methodology to even fewer steps as the technology improves along with our understanding of key scenario elements, i.e., with better, more consistent LLM output one can imagine removing the iterative human checking process between expansion steps (as shown in prior Figure 1). However, even if expansion steps were simplified, we would strongly recommend an HCD approach, beginning and ending with human SMEs to ensure that both initial AI use cases and final expanded scenarios are reflective of real-world AI use. To that end, for any changes in scenario expansion methodology, we envision the need to engage additional SMEs. We plan to expand and replicate with other sectors, such as cybersecurity and manufacturing. Finally, we envision advancing from evaluation scenarios to operationalized metrics and measures. Future efforts will focus on deriving sector-specific measures for human-AI interaction and teaming productivity to support a comprehensive human-centered and sociotechnical assessment.

\section*{Acknowledgements}
The authors would like to sincerely thank the following individuals for their contributions to this work. We extend our gratitude to our colleagues Theodore Jensen and Ryan Steed from the National Institute of Standards and Technology for their insightful technical reviews and invaluable feedback that greatly helped improve this manuscript. We also extend our thanks to Alexander Leblang and Patrick Quentmeyer at the U.S. Department of the Treasury for coordinating with subject matter experts from the financial services sector to review the AI use cases and evaluation scenarios presented in this work. Lastly, we are grateful to Rachel Dzombak at Carnegie Mellon University for facilitating collaboration among the co-authors and supporting the project through essential advocacy.

\section*{Disclaimer}
Any mention of commercial products or reference to commercial organizations is for information only; it does not imply recommendation or endorsement by the National Institute of Standards and Technology (NIST) nor does it imply that the products mentioned are necessarily the best available for the purpose.

\section*{Generative AI Disclosure}
This manuscript was edited with the assistance of Gemini, developed by Google. Gemini was used to refine language, improve clarity, and enhance readability in accordance with the authors’ instructions. All content, scientific claims, and conclusions have been reviewed and verified by the authors to ensure accuracy and originality.

\bibliographystyle{plain}
\bibliography{references}

@misc{aws2025explore,
  title={{Explore AI Use Cases}},
  author={{Amazon Web Services}},
  howpublished={Retrieved December 15, 2025 from \url{https://aws.amazon.com/ai/generative-ai/use-cases/}},
  year={2025}
}

@misc{abecasis2025economy,
  title={{What Is the US Economy’s Potential Growth Rate?}},
  author={Abecasis, Manuel},
  institution={Goldman Sachs Global Investment Research},
  year={2025},
  note={Retrieved December 16, 2025 from \url{https://www.goldmansachs.com/insights/articles/what-is-the-us-economys-potential-growth-rate}}
}

@article{bai2022constitutional,
  title={{Constitutional AI: Harmlessness From AI Feedback}},
  author={Bai, Yuntao and Kadavath, Saurav and Kundu, Sandipan and Askell, Amanda and Kernion, Jackson and Jones, Andy and Chen, Anna and others},
  journal={arXiv preprint arXiv:2212.08073},
  year={2022}
}

@inproceedings{bommasani2024foundation,
  title={{Foundation Model Transparency Reports}},
  author={Bommasani, Rishi and Klyman, Kevin and Longpre, Shayne and Xiong, Betty and Kapoor, Sayash and Maslej, Nestor and Narayanan, Arvind and Liang, Percy},
  booktitle={Proceedings of the AAAI/ACM Conference on AI, Ethics, and Society},
  volume={7},
  pages={181--195},
  year={2024}
}

@misc{canada2025strategy,
  title={{AI Strategy for the Federal Public Service 2025-2027}},
  author={{Government of Canada}},
  howpublished={Retrieved December 16, 2025 from \url{https://www.canada.ca/en/government/system/digital-government/digital-government-innovations/responsible-use-ai/gc-ai-strategy-priority-areas.html}},
  year={2025}
}

@article{chang2024survey,
  title={{A Survey on Evaluation of Large Language Models}},
  author={Chang, Yupeng and Wang, Xu and Wang, Jindong and Wu, Yuan and Yang, Linyi and Zhu, Kaijie and Chen, Hao and Yi, Xiaoyuan and Wang, Cunxiang and Wang, Yidong and others},
  journal={ACM Transactions on Intelligent Systems and Technology},
  volume={15},
  number={3},
  pages={1--45},
  year={2024}
}

@article{chui2018notes,
  title={{Notes From the AI Frontier: Insights From Hundreds of Use Cases}},
  author={Chui, Michael and Manyika, James and Miremadi, Mehdi and Henke, Nicolaus and Chung, Rita and Nel, Pieter and Malhotra, Sankalp},
  journal={McKinsey Global Institute},
  volume={2},
  number={267},
  pages={1--31},
  year={2018},
  note={Retrieved December 15, 2025 from \url{https://www.mckinsey.com/~/media/mckinsey/featured%20insights/artificial%20intelligence/notes%20from%20the%20ai%20frontier%20applications%20and%20value%20of%20deep%20learning/notes-from-the-ai-frontier-insights-from-hundreds-of-use-cases-discussion-paper.pdf}}
}

@article{cui2025effects,
  title={{The Effects of Generative AI on High-Skilled Work: Evidence From Three Field Experiments With Software Developers}},
  author={Cui, Kevin Zheyuan and Demirer, Mert and Jaffe, Sonia and Musolff, Leon and Peng, Sida and Salz, Tobias},
  journal={SSRN},
  year={2025},
  note={Available at SSRN: \url{http://dx.doi.org/10.2139/ssrn.4945566}}
}

@misc{finAIrmf2026framework,
  title={{Financial Services AI Risk Management Framework (FS AI RMF)}},
  author={{Cyber Risk Institute}},
  howpublished={Retrieved April 20, 2026 from \url{https://cyberriskinstitute.org/artificial-intelligence-risk-management/}},
  year={2026}
}

@misc {deloitte2025dossier,
  title={{The AI Dossier: 80+ AI Use Cases: A Collection of New, High-Impact AI Use Cases Organized by Industry, Enterprise Function, and AI Type}},
  author={{Deloitte AI Institute}},
  institution={Deloitte},
  year={2025},
  note={Retrieved December 15, 2025 from \url{https://www.deloitte.com/us/en/what-we-do/capabilities/applied-artificial-intelligence/content/ai-use-cases.html}}
}

@misc {deloitte2025nextgen,
  title={{Next-Gen Controllership: AI and Emerging Tech’s Impact on Finance}},
  author={{Deloitte Center for Controllership}},
  institution={Deloitte},
  year={2025},
  note={Retrieved December 16, 2025 from \url{https://www.deloitte.com/content/dam/assets-zone3/us/en/docs/services/consulting/2025/agentic-ai-dbriefs-poll-results-deck.pdf}}
}

@misc{eu2025strategies,
  title={{Commission Launches Two Strategies to Speed Up AI Uptake in European Industry and Science}},
  author={{European Union}},
  howpublished={Retrieved December 16, 2025 from \url{https://ec.europa.eu/commission/presscorner/detail/en/ip_25_2299}},
  year={2025}
}

@misc{fischerlauder2025empowering,
  title={{Empowering Communities: The Impact of Financial Institutions on Economic Growth}},
  author={Fischer-Lauder, Hannah},
  howpublished={Retrieved Jan.06, 2025 from \url{https://impakter.com/empowering-communities-the-impact-of-financial-institutions-on-economic-growth/}},
  year={2025}
}

@inproceedings{gailmard2025known,
  title={{Known Unknowns and Unknown Unknowns: Designing a Scalable Adverse Event Reporting System for AI}},
  author={Gailmard, Lindsey and Spence, Drew and Lawrence, Christie and Ho, Daniel E.},
  booktitle={Proceedings of the AAAI/ACM Conference on AI, Ethics, and Society},
  volume={8},
  pages={1004--1017},
  year={2025}
}

@article{ganguli2022red,
  title={{Red Teaming Language Models to Reduce Harms: Methods, Scaling Behaviors, and Lessons Learned}},
  author={Ganguli, Deep and Lovitt, Liane and Kernion, Jackson and Askell, Amanda and Bai, Yuntao and Kadavath, Saurav Bull and Mann, Ben and others},
  journal={arXiv preprint arXiv:2209.07858},
  year={2022}
}

@misc{google2025usecases,
  title={{1,001 Real-World Gen AI Use Cases From the World's Leading Organizations}},
  author={{Google}},
  howpublished={Retrieved December 15, 2025 from \url{https://cloud.google.com/transform/101-real-world-generative-ai-use-cases-from-industry-leaders}},
  year={2025}
}

@misc{jhu2025symposium,
  title={{Human + AI: Redefining the Standard of Care in Medicine}},
  author={{Johns Hopkins, Malone Center for Engineering in Healthcare}},
  howpublished={The 9th Annual Johns Hopkins Research Symposium on Engineering in Healthcare. \url{https://malonecenter.jhu.edu/johns-hopkins-malone-center-2025-symposium/}},
  year={2025}
}

@misc{hoyek2024promise,
  title={{Gen AI’s Productivity Promise: Huge Potential but Most Have Not Yet Reached Scaled Impact}},
  author={El Hoyek, Marie and M\"{u}ller, Nicolai and Ronellenfitsch, Jonas},
  howpublished={McKinsey \& Company. Retrieved December 16, 2025 from \url{https://www.mckinsey.com/capabilities/operations/our-insights/operations-blog/gen-ais-productivity-promise-huge-potential-but-most-have-not-yet-reached-scaled-impact}},
  year={2024}
}

@misc{india2025transforming,
  title={{Transforming India With AI}},
  author={{Government of India}},
  howpublished={Retrieved December 16, 2025 from \url{https://www.pib.gov.in/PressReleasePage.aspx?PRID=2178092&reg=3&lang=2}},
  year={2025}
}

@misc{iso9241-11,
  title={{ISO 9241-11:2018 Ergonomics of Human-System Interaction --- Part 11: Usability: Definitions and Concepts}},
  author={{ISO}},
  institution={International Organization for Standardization},
  year={2018},
  url={https://www.iso.org/standard/63500.html}
}

@misc{iso9241-210,
  title={{ISO 9241-210:2019 Ergonomics of Human-System Interaction --- Part 210: Human-Centred Design for Interactive Systems}},
  author={{ISO}},
  institution={International Organization for Standardization},
  year={2019},
  url={https://www.iso.org/standard/77520.html}
}

@misc{iso42001,
  title={{ISO/IEC 42001:2023 Information Technology --- Artificial Intelligence --- Management System}},
  author={{ISO/IEC}},
  institution={International Organization for Standardization},
  year={2023},
  url={https://www.iso.org/standard/42001}
}

@misc{iso24030,
  title={{ISO/IEC TR 24030:2024 Information Technology – Artificial Intelligence (AI) – Use Cases}},
  author={{ISO/IEC}},
  institution={International Organization for Standardization},
  year={2024},
  url={https://www.iso.org/standard/84144.html}
}

@misc{japan2025innovation,
  title={{Integrated Innovation Strategy 2025}},
  author={{Cabinet Office, Government of Japan}},
  howpublished={Retrieved December 16, 2025 from \url{https://www8.cao.go.jp/cstp/tougosenryaku/togo2025_honbun_eiyaku.pdf}},
  year={2025}
}

@article{karanxha2025evaluating,
  title={{Evaluating Transparency in the Development of Artificial Intelligence Systems: A Systematic Literature Review}},
  author={Karanxha, Giulia and Ofem, Paulinus},
  journal={International Journal of Advanced Computer Science \& Applications},
  volume={16},
  number={10},
  year={2025}
}

@misc{kpmg2024finance,
  title={{KPMG Global AI in Finance Report}},
  author={{KPMG}},
  institution={KPMG},
  year={2024},
  note={Retrieved December 16, 2025 from \url{https://assets.kpmg.com/content/dam/kpmgsites/xx/pdf/2024/11/ai-in-finance.pdf.coredownload.inline.pdf}}
}

@inproceedings{lee2025impact,
  title={{The Impact of Generative AI on Critical Thinking: Self-Reported Reductions in Cognitive Effort and Confidence Effects From a Survey of Knowledge Workers}},
  author={Lee, Hao-Ping and Sarkar, Advait and Tankelevitch, Lev and Drosos, Ian and Rintel, Sean and Banks, Richard and Wilson, Nicholas},
  booktitle={Proceedings of the 2025 CHI Conference on Human Factors in Computing Systems},
  pages={1--22},
  year={2025}
}

@article{liang2022holistic,
  title={{Holistic Evaluation of Language Models}},
  author={Liang, Percy and Bommasani, Rishi and Lee, Tony and Tsipras, Dimitris and Soylu, Dilara and Yasunaga, Michihiro and Zhang, Yian and others},
  journal={arXiv preprint arXiv:2211.09110},
  year={2022}
}

@article{maslej2025index,
  title={{The AI Index 2025 Annual Report}},
  author={Maslej, Nestor and Fattorini, Loredana and Perrault, Raymond and Gil, Yolanda and Parli, Vanessa and Kariuki, Njenga and Capstick, Emily and others},
  journal={arXiv preprint arXiv:2504.07139},
  year={2025},
  note={AI Index Steering Committee, Institute for Human-Centered AI, Stanford University}
}

@article{mazeika2024harmbench,
  title={{Harmbench: A Standardized Evaluation Framework for Automated Red Teaming and Robust Refusal}},
  author={Mazeika, Mantas and Phan, Long and Yin, Xuwang and Zou, Andy and Wang, Zifan and Mu, Norman and Sakhaee, Elham and others},
  journal={arXiv preprint arXiv:2402.04249},
  year={2024}
}

@article{milley2025markets,
  title={{Financial Markets: The Backbone of the Global Economy}},
  author={Milley, Anna},
  journal={Journal of Stock \& Forex Trading},
  volume={12},
  pages={292},
  year={2025}
}

@misc{murray2025labor,
  title={{How Artificial Intelligence Impacts the US Labor Market}},
  author={Murray, Seb},
  howpublished={MIT Sloan School of Management. Retrieved December 16, 2025 from \url{https://mitsloan.mit.edu/ideas-made-to-matter/how-artificial-intelligence-impacts-us-labor-market}},
  year={2025}
}

@inproceedings{naidu2023review,
  title={{A Review of Evaluation Metrics in Machine Learning Algorithms}},
  author={Naidu, Gireen and Zuva, Tranos and Sibanda, Elias Mmbongeni},
  booktitle={Computer Science On-line Conference},
  pages={15--25},
  publisher={Springer International Publishing},
  year={2023}
}

@misc{nist2023rmf,
  title={{Artificial Intelligence Risk Management Framework (AI RMF 1.0)}},
  author={{NIST AI 100-1}},
  institution={National Institute of Standards and Technology},
  year={2023},
  doi={10.6028/NIST.AI.100-1}
}

@misc{nist2024genai,
  title={{Artificial Intelligence Risk Management Framework: Generative Artificial Intelligence Profile}},
  author={{NIST AI 600-1}},
  institution={National Institute of Standards and Technology},
  year={2024},
  doi={10.6028/NIST.AI.600-1}
}

@misc{nist2025aria,
  title={{Assessing Risks and Impacts of AI (ARIA): Pilot Evaluation Report}},
  author={{NIST AI 700-2}},
  institution={National Institute of Standards and Technology},
  year={2025},
  doi={10.6028/NIST.AI.700-2}
}

@misc{oecd2025principles,
  title={{OECD AI Principles}},
  author={{OECD}},
  howpublished={Retrieved December 16, 2025 from \url{https://oecd.ai/en/ai-principles}},
  year={2025}
}

@misc{oecd2025catalogue,
  title={{Catalogue of Tools \& Metrics for Trustworthy AI, List of Metric Use Cases}},
  author={{OECD.AI}},
  howpublished={Retrieved December 16, 2025 from \url{https://oecd.ai/en/catalogue/metric-use-cases}},
  year={2025}
}

@misc{olmstead2025transformation,
  title={{Digital Transformations \& Tech Adoption by Sector (2025)}},
  author={Olmstead, Levi},
  howpublished={Retrieved January 06, 2026 from \url{https://whatfix.com/blog/digital-transformation-by-sector/}},
  year={2025}
}

@article{powers2011evaluation,
  title={{Evaluation: From Precision, Recall and F-measure to ROC, Informedness, Markedness \& Correlation}},
  author={Powers, David},
  journal={Journal of Machine Learning Technologies},
  volume={2},
  number={1},
  pages={37--63},
  year={2011}
}

@inproceedings{pushkarna2022datacards,
  title={{Data Cards: Purposeful and Transparent Dataset Documentation for Responsible AI}},
  author={Pushkarna, Mahima and Zaldivar, Andrew and Kjartansson, Oddur},
  booktitle={Proceedings of the 2022 ACM Conference on Fairness, Accountability, and Transparency},
  pages={1776--1826},
  year={2022}
}

@inproceedings{radiyadixit2023audit,
  title={{A Sociotechnical Audit: Assessing Police Use of Facial Recognition}},
  author={Radiya-Dixit, Evani and Neff, Gina},
  booktitle={Proceedings of the 2023 ACM Conference on Fairness, Accountability, and Transparency},
  pages={1334--1346},
  year={2023}
}

@misc{roose2025measurement,
  title={{A.I. Has a Measurement Problem}},
  author={Roose, Kevin},
  howpublished={The New York Times. Retrieved January 6, 2026 from \url{https://www.nytimes.com/2024/04/15/technology/ai-models-measurement.html}},
  year={2025}
}

@inproceedings{schroeder2025disclosure,
  title={{Disclosure Without Engagement: An Empirical Review of Positionality Statements at FAccT}},
  author={Schroeder, Hope and Pareek, Akshansh and Barocas, Solon},
  booktitle={Proceedings of the 2025 ACM Conference on Fairness, Accountability, and Transparency},
  pages={1195--1210},
  year={2025}
}

@inproceedings{selbst2019fairness,
  title={{Fairness and Abstraction in Sociotechnical Systems}},
  author={Selbst, Andrew D. and Boyd, danah and Friedler, Sorelle A. and Venkatasubramanian, Suresh and Vertesi, Janet},
  booktitle={Proceedings of the 2019 ACM Conference on Fairness, Accountability, and Transparency},
  pages={59--68},
  year={2019}
}

@inproceedings{sun2024wild,
  title={{Generative AI in the Wild: Prospects, Challenges, and Strategies}},
  author={Sun, Yuan and Jang, Eunchae and Ma, Fenglong and Wang, Ting},
  booktitle={Proceedings of the 2024 CHI Conference on Human Factors in Computing Systems},
  pages={1--16},
  year={2024}
}

@misc{census2025btos,
  title={{Business Trends and Outlook Survey (BTOS) Key Performance Indicators}},
  author={{U.S. Census}},
  howpublished={Retrieved December 16, 2025 from \url{https://www.census.gov/hfp/btos/data}},
  year={2025}
}

@misc{chamber2025smallbusiness,
  title={{Empowering Small Business: The Impact of Technology on U.S. Small Business, 4th Ed.}},
  author={{U.S. Chamber of Commerce}},
  year={2025},
  note={Retrieved December 16, 2025 from \url{https://www.uschamber.com/assets/documents/20251621-CTEC-Empowering-Small-Business-Report-2025-v1-r10-Digital-FINAL.pdf}}
}

@misc{cioc2024inventory,
  title={{2024 Federal AI Use Case Inventory}},
  author={{U.S. Chief Information Officers Council}},
  howpublished={Retrieved December 15, 2025 from \url{https://www.cio.gov/policies-and-priorities/Executive-Order-13960-AI-Use-Case-Inventories-Reference/}},
  year={2024}
}

@misc {whitehouse2025actionplan,
  title={{Winning the Race: America's AI Action Plan}},
  author={{U.S. The White House}},
  year={2025},
  note={Retrieved December 16, 2025 from \url{https://www.whitehouse.gov/wp-content/uploads/2025/07/Americas-AI-Action-Plan.pdf}}
}

@misc {treasury2024financial,
  title={{Artificial Intelligence in Financial Services: Report on the Uses, Opportunities, and Risks of Artificial Intelligence in the Financial Services Sector}},
  author={{U.S. Treasury}},
  year={2024},
  note={Retrieved December 16, 2025 from \url{https://home.treasury.gov/system/files/136/Artificial-Intelligence-in-Financial-Services.pdf}}
}

@misc {gao2025genai,
  title={{Generative {AI} Use and Management at Federal Agencies}},
  author={{U.S. Government Accountability Office}},
  year={2025},
  note={Accessed April 27, 2026 from \url{https://www.gao.gov/products/gao-25-107653}}
}

@inproceedings{vannostrand2024actionable,
  title={{Actionable Recourse for Automated Decisions: Examining the Effects of Counterfactual Explanation Type and Presentation on Lay User Understanding}},
  author={VanNostrand, Peter M. and Hofmann, Dennis M. and Ma, Lei and Rundensteiner, Elke A.},
  booktitle={Proceedings of the 2024 ACM Conference on Fairness, Accountability, and Transparency},
  pages={1682--1700},
  year={2024}
}

@article{verga2024judges,
  title={{Replacing Judges With Juries: Evaluating LLM Generations With a Panel of Diverse Models}},
  author={Verga, Pat and Hofstatter, Sebastian and Althammer, Sophia and Su, Yixuan and Piktus, Aleksandra and Arkhangorodsky, Arkady and Xu, Minjie and White, Naomi and Lewis, Patrick},
  journal={arXiv preprint arXiv:2404.18796},
  year={2024}
}

@article{walker2008sociotechnical,
  title={{A Review of Sociotechnical Systems Theory: A Classic Concept for New Command and Control Paradigms}},
  author={Walker, Guy H. and Stanton, Neville A. and Salmon, Paul M. and Jenkins, Daniel P.},
  journal={Theoretical Issues in Ergonomics Science},
  volume={9},
  number={6},
  pages={479--499},
  year={2008}
}

@article{wallach2025measurement,
  title={{Position: Evaluating Generative AI Systems Is a Social Science Measurement Challenge}},
  author={Wallach, Hanna and Desai, Meera and Cooper, A. Feder and Wang, Angelina and Atalla, Chad and Barocas, Solon and Blodgett, Su Lin and others},
  journal={arXiv preprint arXiv:2502.00561},
  year={2025}
}

@inproceedings{wei2023jailbroken,
  title={{Jailbroken: How Does LLM Safety Training Fail?}},
  author={Wei, Alexander and Haghtalab, Nika and Steinhardt, Jacob},
  booktitle={Advances in Neural Information Processing Systems},
  volume={36},
  pages={80079--80110},
  year={2023}
}

@inproceedings{yue2024mmmu,
  title={{MMMU: A Massive Multi-Discipline Multimodal Understanding and Reasoning Benchmark for Expert AGI}},
  author={Yue, Xiang and Ni, Yuansheng and Kai Zhang and Zheng, Tianyu and Liu, Ruoqi and Zhang, Ge and Stevens, Samuel and others},
  booktitle={Proceedings of the IEEE/CVF Conference on Computer Vision and Pattern Recognition},
  pages={9556--9567},
  year={2024}
}

@inproceedings{zeng2025airbench,
  title={{AIR-Bench 2024: A Safety Benchmark Based on Regulation and Policies Specified Risk Categories}},
  author={Zeng, Yi and Yang, Yu and Zou, Andy and Tan, Jeffrey Ziwei and Tu, Yuheng and Mai, Yifan and Klyman, Kevin and Pan, Minzhou and Jia, Ruoxi and Song, Dawn and Liang, Percy and Li, Bo},
  booktitle={The Thirteenth International Conference on Learning Representations},
  year={2025}
}

@inproceedings{zheng2023judging,
  title={{Judging LLM-as-a-Judge With MT-bench and Chatbot Arena}},
  author={Zheng, Lianmin and Chiang, Wei-Lin and Sheng, Ying and Zhuang, Siyuan and Wu, Zhanghao and Zhuang, Yonghao and Lin, Zi and others},
  booktitle={Advances in Neural Information Processing Systems},
  volume={36},
  pages={46595--46623},
  year={2023}
}

@article{zou2025security,
  title={{Security Challenges in AIAgent Deployment: Insights From a Large Scale Public Competition}},
  author={Zou, Andy and Lin, Maxwell and Jones, Eliot and Nowak, Micha and Dziemian, Mateusz and Winter, Nick and Grattan, Alexander and others},
  journal={arXiv preprint arXiv:2507.20526},
  year={2025}
}

\clearpage
\begin{landscape}
\section*{Appendix}
As discussed in Section \ref{sec:sec4}, a summary list of the six high-level SME-identified use cases and the 107 scenarios generated is included in Table 4. The table shows the following scenario elements: Use Case, Scenario Title, and Scenario Description. These AI use cases and scenarios provided are illustrative of methodology and by no means exhaustive for the financial services sector. We note that many use cases could be expanded in additional and different ways—for example, “credit memo generation” includes activities such as underwriting, risk rating, stress testing, and loan loss provisioning—but we simplify for the sake of exposition and visual presentation in tabular format.\par


\begin{center}
\begin{longtable}{p{4.5cm}p{5.5cm}p{11cm}}
\caption{Summary of Financial-Sector Use Cases and Scenarios} \label{tab:financial_scenarios} \\
\toprule
\textbf{Use Case} & \textbf{Scenario Title} & \textbf{Scenario Description} \\ \midrule
\endfirsthead

\multicolumn{3}{c}%
{{\tablename\ \thetable{} -- continued from previous page}} \\
\toprule
\textbf{Use Case} & \textbf{Scenario Title} & \textbf{Scenario Description} \\ 
\midrule
\endhead

\midrule
\multicolumn{3}{r}{{\footnotesize{\textit{Continued on next page}}}} \\
\bottomrule
\endfoot

\bottomrule
\endlastfoot
Credit Memo Generation & Automating credit memo generation & AI automatically generates draft credit memos using financial data, borrower profiles, and credit history. \\ 
Credit Memo Generation & Small business automated underwriting & AI generates standardized credit memos for small business loans using alternative data sources and simplified financial metrics. \\ 
Credit Memo Generation & Retail credit decisioning & AI generates consumer credit memos for personal loans, credit cards, and mortgages using behavioral data and credit bureau information. \\ 
Credit Memo Generation & Multi-entity credit analysis & AI generates consolidated credit memos for complex corporate structures with multiple subsidiaries and guarantors. \\ 
Credit Memo Generation & Trade finance credit documentation & AI creates credit memos for trade finance transactions, incorporating country risk, commodity pricing, and supply chain factors. \\ 
Credit Memo Generation & Cross-border credit evaluation & AI creates credit memos for international lending by incorporating foreign exchange risk, political risk, and cross-jurisdictional regulatory requirements. \\ 
Credit Memo Generation & Covenant monitoring and waivers & AI monitors loan covenants and generates credit memos for covenant violations and waiver requests. \\ 
Credit Memo Generation & Stress testing and scenario analysis & AI conducts credit stress testing and generates memos analyzing portfolio performance under adverse scenarios. \\ 
Credit Memo Generation & Loan loss provision automation & AI automates loan loss provision calculations and credit memo generation for impaired assets. \\ 
Credit Memo Generation & Credit risk rating adjustments & AI recommends credit risk rating changes and generates supporting credit memos. \\ 
Credit Memo Generation & Credit concentration management & AI analyzes credit concentrations and generates memos for concentration risk management. \\ 
Credit Memo Generation & Industry and sector analysis & AI conducts industry analysis and generates credit memos for sector-specific credit risks. \\ 
Credit Memo Generation & ESG-integrated credit assessment & AI incorporates environmental, social, and governance factors into traditional credit memos for sustainable finance initiatives. \\ 
Credit Memo Generation & Collateral valuation analysis & AI analyzes collateral values and generates credit memos for collateral-dependent loans. \\ 
Credit Memo Generation & Regulatory capital allocation & AI calculates regulatory capital requirements and generates memos for credit risk-weighted assets. \\ 
Credit Memo Generation & Credit decision documentation & AI generates comprehensive documentation for credit committee decisions and approvals. \\ 
Cyber Defense Enablement & Cloud security configuration & AI manages cloud security configurations and compliance for financial institution cloud deployments. \\ 
Cyber Defense Enablement & Data loss prevention optimization & AI optimizes data loss prevention (DLP) systems to protect sensitive financial information. \\ 
Cyber Defense Enablement & Incident response automation & AI automates cybersecurity incident response and containment procedures. \\ 
Cyber Defense Enablement & Security metric analysis and reporting & AI analyzes security metrics across multiple tools and platforms to provide executive-level security posture reporting and trend analysis. \\ 
Cyber Defense Enablement & Regulatory cybersecurity reporting & AI automates cybersecurity regulatory reporting and compliance documentation. \\ 
Cyber Defense Enablement & Network segmentation analysis & AI analyzes network architecture and recommends security segmentation strategies. \\ 
Cyber Defense Enablement & Security awareness training optimization & AI optimizes cybersecurity awareness training programs for financial institution employees. \\ 
Cyber Defense Enablement & Threat intelligence correlation & AI correlates internal security events with external threat intelligence feeds to provide contextual awareness and attribution insights. \\ 
Cyber Defense Enablement & Malware analysis and sandboxing & AI automatically analyzes suspicious files and URLs in isolated environments to determine maliciousness and extraction of indicators of compromise. \\ 
Cyber Defense Enablement & Insider threat behavioral analysis & AI analyzes employee behavior patterns to identify potential insider cybersecurity threats. \\ 
Cyber Defense Enablement & Phishing email triage & Use AI to triage suspicious emails, flag likely phishing attempts, and reduce analyst burden by prioritizing critical threats. \\ 
Cyber Defense Enablement & Network anomaly detection & AI monitors network traffic patterns to identify suspicious activities, lateral movement, and potential data exfiltration attempts in real-time. \\ 
Cyber Defense Enablement & Identity and access anomaly detection & AI monitors user authentication patterns and access behaviors to detect compromised accounts and insider threats. \\ 
Cyber Defense Enablement & Threat intelligence analysis & AI analyzes cyber threat intelligence to identify relevant threats to financial institutions. \\ 
Cyber Defense Enablement & Due diligence and monitoring of third-party service providers & AI assesses cybersecurity risks from third-party vendors and service providers. \\ 
Cyber Defense Enablement & Vulnerability management prioritization & AI prioritizes cybersecurity vulnerabilities for remediation based on risk assessment. \\ 
Developer Productivity & Code completion & Enhancing developer productivity with AI tools. \\ 
Developer Productivity & Code review automation & AI analyzes code changes and provides automated review feedback based on internal standards. \\ 
Developer Productivity & Code refactoring suggestions & AI identifies code that needs refactoring and suggests improvements for maintainability and performance. \\ 
Developer Productivity & Commit message generation & AI helps generate meaningful commit messages based on code diffs. \\ 
Developer Productivity & Code documentation generation & AI automatically generates technical documentation from codebases and comments. \\ 
Developer Productivity & Unit test generation & AI creates comprehensive unit tests based on code analysis and edge case identification. \\ 
Developer Productivity & Integration test generation & Automatically generate integration tests for components that interact, based on code and service dependencies. \\ 
Developer Productivity & Automated testing strategy & AI recommends testing strategies and generates test automation frameworks. \\ 
Developer Productivity & Environment configuration management & AI manages configuration across development, staging, and production environments. \\ 
Developer Productivity & Automated code deployment & AI assists in automating deployment pipelines and release management. \\ 
Developer Productivity & Database migration scripting & AI generates database migration scripts for schema changes and data transformations. \\ 
Developer Productivity & Error log analysis and debugging & AI analyzes application logs to identify patterns, suggest fixes, and prioritize debugging efforts. \\ 
Developer Productivity & Performance profiling analysis & AI analyzes application performance profiles and suggests optimization strategies. \\ 
Developer Productivity & Performance monitoring integration & AI helps integrate and configure application performance monitoring solutions. \\ 
Developer Productivity & Dependency vulnerability scanning & AI scans code dependencies and identifies security vulnerabilities in third-party libraries. \\ 
Developer Productivity & Security patch recommendation & AI analyzes codebases to recommend security patches and vulnerability fixes. \\ 
Developer Productivity & Dependency management optimization & AI manages and optimizes software dependencies and library versions. \\ 
Developer Productivity & API documentation generation & Automatically generate API documentation from code annotations and usage patterns. \\ 
Developer Productivity & API rate limiting configuration & AI helps configure API rate limiting and throttling policies for financial services. \\ 
Developer Productivity & Architecture diagram generation & AI creates system architecture diagrams from source code and config files to visualize structure and dependencies. \\ 
Developer Productivity & Database query optimization & AI analyzes and optimizes SQL queries for performance and security best practices. \\ 
Financial Crimes Aggregation & AI aggregates and summarizes AML issues and fraud cases & AI supports financial crime investigation by aggregating and summarizing AML issues and fraud cases. \\ 
Financial Crimes Aggregation & Pattern recognition in suspicious transactions & AI identifies patterns and correlations across multiple financial crime cases to detect emerging threats. \\ 
Financial Crimes Aggregation & Temporal analysis and trend identification & AI analyzes case data over time to identify trends and evolving criminal methodologies. \\ 
Financial Crimes Aggregation & Case outcome prediction & AI predicts likely outcomes for ongoing financial crime investigations. \\ 
Financial Crimes Aggregation & Evidence synthesis and summarization & AI synthesizes evidence from multiple sources into comprehensive case summaries. \\ 
Financial Crimes Aggregation & Asset tracing and recovery support & AI assists in tracing assets and identifying recovery opportunities in financial crime cases. \\ 
Financial Crimes Aggregation & Customer risk assessment automation & AI automates customer risk assessments and periodic reviews for AML compliance programs. \\ 
Financial Crimes Aggregation & Beneficial ownership analysis & AI analyzes complex ownership structures to identify ultimate beneficial owners and hidden relationships. \\ 
Financial Crimes Aggregation & Regulatory reporting automation & AI generates standardized reports for regulatory submissions based on aggregated case data. \\ 
Financial Crimes Aggregation & Cryptocurrency transaction monitoring & AI monitors and analyzes cryptocurrency transactions for money laundering and illicit activities. \\ 
Financial Crimes Aggregation & Cross-border transaction analysis & AI analyzes international wire transfers and cross-border payments for suspicious patterns. \\ 
Financial Crimes Aggregation & Trade-based money laundering detection & AI identifies suspicious trade finance activities and invoice manipulation schemes. \\ 
Financial Crimes Aggregation & Sanctions screening automation & AI automates sanctions list screening and identifies potential violations across customer transactions. \\ 
Financial Crimes Aggregation & Insider threat detection & AI monitors employee activities and transactions to identify potential insider threats and misconduct. \\ 
Financial Crimes Aggregation & Regulatory examination preparation & AI assists in preparing comprehensive responses for regulatory examinations and audits. \\ 
Financial Crimes Aggregation & Suspicious activity reporting optimization & AI optimizes the quality and timing of suspicious activity report (SAR) filings. \\ 
Financial Crimes Aggregation & Financial intelligence sharing & AI facilitates information sharing with law enforcement and financial intelligence units. \\ 
Internal Call Center Support & Real-time agent coaching & AI provides real-time suggestions and coaching to agents during customer calls. \\ 
Internal Call Center Support & Call outcome prediction & AI predicts likely call outcomes and resolution times based on initial customer interactions. \\ 
Internal Call Center Support & Knowledge base query assistance & AI helps agents quickly find relevant information from internal knowledge bases during calls. \\ 
Internal Call Center Support & Call quality scoring & AI evaluates and scores call center interactions for quality assurance. \\ 
Internal Call Center Support & Call recording and transcription analysis & AI analyzes call recordings and transcriptions for insights and compliance. \\ 
Internal Call Center Support & Customer sentiment analysis & AI analyzes customer call transcripts to identify sentiment patterns and escalation risks. \\ 
Internal Call Center Support & Escalation decision support & AI recommends when to escalate customer issues to specialized departments. \\ 
Internal Call Center Support & Crisis communication management & AI assists agents in managing communications during financial institution crises. \\ 
Internal Call Center Support & Automated call routing & AI routes incoming calls to appropriate agents based on inquiry type and agent expertise. \\ 
Internal Call Center Support & Customer callback scheduling & AI optimizes callback scheduling based on customer preferences and agent availability. \\ 
Internal Call Center Support & Multilingual support optimization & AI assists with language detection and routes calls to appropriate multilingual agents. \\ 
Internal Call Center Support & Customer data access control & AI manages agent access to customer financial information during support calls. \\ 
Internal Call Center Support & Customer authentication verification & AI assists in verifying customer identity during phone support interactions. \\ 
Internal Call Center Support & Regulatory compliance monitoring & AI monitors call center interactions for regulatory compliance violations. \\ 
Internal Call Center Support & Complaint categorization and tracking & AI automatically categorizes customer complaints and tracks resolution progress. \\ 
Internal Call Center Support & Transaction dispute resolution & AI assists agents in resolving customer transaction disputes and chargebacks. \\ 
Internal Call Center Support & Agent performance analytics & AI analyzes agent performance metrics and provides insights for coaching and development. \\ 
Internal Call Center Support & Customer retention decision support & AI provides recommendations for customer retention strategies during service calls. \\ 
Internal Call Center Support & Cross-selling opportunity identification & AI identifies appropriate cross-selling opportunities during customer service calls. \\ 
SAR Filing & Cross-system data integration and correlation & AI integrates and correlates data from multiple banking systems, databases, and external sources to provide comprehensive information for SAR preparation and filing. \\ 
SAR Filing & Geographic and temporal correlation & AI identifies geographic and temporal patterns in suspicious activities for enhanced SAR context. \\ 
SAR Filing & Law enforcement feedback integration & AI incorporates law enforcement feedback to improve SAR filing quality and relevance. \\ 
SAR Filing & Cross-institutional coordination & AI coordinates SAR filing with other financial institutions to avoid duplicate reporting. \\ 
SAR Filing & Narrative generation assistance & AI helps generate comprehensive narratives for SAR filings based on available case data. \\ 
SAR Filing & Supporting documentation compilation & AI assists in gathering and organizing supporting documentation for SAR filings. \\ 
SAR Filing & Narrative quality enhancement & AI improves SAR narrative quality and completeness for law enforcement utility. \\ 
SAR Filing & Subject identification and research & AI assists in identifying and researching SAR subjects for comprehensive reporting. \\ 
SAR Filing & Streamlining SAR filing & AI assists compliance teams by streamlining SAR filing. \\ 
SAR Filing & Threshold monitoring and alerting & AI monitors transaction patterns against SAR filing thresholds and generates alerts for potential violations. \\ 
SAR Filing & Risk scoring and case prioritization & AI evaluates and scores suspicious activity cases based on risk factors, transaction amounts, customer profiles, and regulatory requirements to prioritize SAR preparation and filing order. \\ 
SAR Filing & AI determines when activities meet the threshold for SAR filing requirements & AI analyzes transaction data and customer behavior to identify complex suspicious patterns that may require SAR filing, including layering, structuring, and other money laundering typologies. \\ 
SAR Filing & Suspicious activity threshold determination & AI determines when activities meet the threshold for SAR filing requirements. \\ 
SAR Filing & Filing deadline management & AI manages SAR filing deadlines and prioritizes urgent filings for timely submission. \\ 
SAR Filing & Continuing activity determination & AI determines when to file continuing activity SARs for ongoing suspicious patterns. \\ 
SAR Filing & Regulatory template compliance verification & AI ensures SAR filings conform to current regulatory templates, field requirements, and formatting standards across different jurisdictions and reporting requirements. \\ 
SAR Filing & Regulatory guidance interpretation & AI interprets current regulatory guidance and applies it to SAR filing decisions. \\ 
SAR Filing & Quality assurance and review & AI conducts comprehensive quality assurance reviews of SAR filings before submission. \\
\end{longtable}
\end{center}

\end{landscape}
\end{document}